\documentclass[final,5p,times]{elsarticle}

\usepackage[utf8]{inputenc}
\usepackage[T1]{fontenc}
\usepackage[english]{babel}

\usepackage{microtype}
\usepackage{amssymb}
\usepackage{amsmath}
\usepackage{amsfonts}
\usepackage{graphicx}
\usepackage{bm}

\usepackage[footnotesize]{caption}
\usepackage{xurl}
\usepackage{hyperref}

\journal{Physics Letters B}

\begin{document}
	
	\begin{frontmatter}
		
	\title{Role of pre-equilibrium particle emission in the
		$^{249-251}$No production in the
		$^{204}$Pb($^{48}$Ca,~xn)$^{252-x}$No reaction}
			
		\author[first]{Yu.L.~Parfenova}
		\ead{parfenova@jinr.ru}
		
		\author[first,second]{M.S.~Tezekbayeva}
		
		\author[first]{I.A.~Egorova}
		
		\author[first,second]{A.N.~Ismailova}

		\affiliation[first]{
			organization={Flerov Laboratory of Nuclear Reactions, Joint Institute
				for Nuclear Research},
			addressline={},
			city={Dubna},
			postcode={141980},
			state={},
			country={Russia}
		}
		
		\affiliation[second]{
			organization={Institute of Nuclear Physics},
			addressline={},
			city={Almaty},
			postcode={050032},
			state={},
			country={Republic of Kazakhstan}
		}

\begin{abstract}
	
	Experimental data on the production of $^{250}$No in its ground and
	isomeric states in the $^{204}$Pb($^{48}$Ca,2n)$^{250}$No reaction at
	$^{48}$Ca beam energies from 200 to 240 MeV are analyzed within a unified
	approach combining the Hauser--Feshbach statistical model and the Griffin
	exciton model, assuming the emission of one pre-equilibrium particle.
	The influence of pre-equilibrium processes on the production cross
	sections of the $^{249-251}$No isotopes is investigated.
	The inclusion of pre-equilibrium light-particle emission significantly
	improves the description of the measured production cross sections,
	particularly in the high-energy region, where differences of several
	orders of magnitude are obtained, and for the population of the ground
	and isomeric states of $^{250}$No.
	The proposed approach also reproduces the observed incident-energy
	dependence of the residual-nucleus production cross sections.
	
\end{abstract}

\begin{keyword}
	pre-equilibrium emission \sep Hauser--Feshbach model \sep exciton model \sep
	nobelium isotopes \sep isomeric states \sep fusion--evaporation reactions
\end{keyword}

\end{frontmatter}
	
\section{Introduction}
\label{introduction}
%===============================================================================

The investigation of heavy and superheavy nuclei is an important area of research aimed at exploring the limits of nuclear stability. Over the past three decades, substantial experimental and theoretical progress has been achieved in the synthesis and study of superheavy elements, significantly extending the chart of nuclides and providing new information on the properties of nuclei at the limits of stability. Spectroscopic studies of heavy-ion-induced reactions provide valuable information on the structure of short-lived superheavy nuclei near the boundary of nuclear stability \cite{ACKERMANN2026104215,Dracoulis:2016koy}. Nuclei far from the stability line exhibit characteristic features associated with shell closures, unusual collective modes, and decay properties.

Fusion--evaporation reactions induced by heavy ions constitute one of the principal methods for producing heavy and superheavy nuclei. A reliable theoretical description of these reactions is therefore important for planning experiments and interpreting the measured production cross sections and decay properties.

Superheavy nuclei produced in complete-fusion reactions of heavy-ion beams with heavy target nuclei, followed by neutron evaporation from the excited compound nucleus, often have short half-lives and small production cross sections. These nuclei are identified using the correlation-analysis method, which links recoil-implantation events with subsequent $\alpha$-decay and spontaneous-fission events on the basis of their energies, time intervals, and positions in the focal-plane detector \cite{SHofmann_1998,Tezekbayeva2021}. The resulting genetically correlated decay chains allow the produced nuclei to be identified and their production cross sections to be determined.

Recently, the $^{204}$Pb($^{48}$Ca,2n)$^{250}$No reaction leading to the production of $^{250}$No in the ground ($0^+$) and isomeric ($6^+$) states, denoted as $^{250g,m}$No, has been investigated in detail at laboratory energies from 200 to 250 MeV \cite{Tezekbayeva2021,Tezekbayeva2022,Mukhin:2024vfi} using the experimental facilities of the Flerov Laboratory of Nuclear Reactions, Joint Institute for Nuclear Research (FLNR JINR, Dubna, Russia). Detailed experimental data were obtained for the production of the $^{249-251}$No isotopes, including the production cross sections of $^{250g,m}$No and the corresponding isomeric ratios.

K isomerism is of particular interest because of its characteristic transition-selection rules and hindrance effects. The K-isomeric states are associated with high-spin configurations, and their population in nuclear reactions depends on the angular-momentum distribution and on the population of low-lying states in the produced nuclei. In heavy-ion-induced reactions, relatively large angular momentum is transferred to the compound nucleus, allowing the K-isomeric states to be populated efficiently even at near-barrier energies. Measurements of the isomeric-state production can therefore provide information on angular-momentum distributions and the underlying reaction mechanisms.

A statistical-model description was previously applied to the experimental data on the nobelium-isotope production reported in Ref.~\cite{Tezekbayeva2022}. The calculations reproduced the experimental production cross sections at lower incident energies and underestimated the data in the high-energy region. This discrepancy suggests that additional reaction mechanisms, in particular pre-equilibrium particle emission, may contribute to the production of the residual nuclei.

The production of nuclei in isomeric states and the sensitivity of isomeric ratios to angular-momentum distributions attracted considerable attention in earlier studies \cite{DiGregorio:1990pv,Matsuo:1965pr,Qaim:1988zz,Glebov1991,Swanson:1970yxs,Tulinov1993,Grudzevich1994,Canto:2020}. In particular, the angular-momentum distributions of rare-earth nuclei produced in $\alpha$-particle-induced reactions and the role of pre-equilibrium emission in the population of isomeric states were investigated within statistical models of nuclear reactions \cite{Tulinov1993}.

At incident energies of a few MeV per nucleon, the de-excitation of the compound nucleus may involve both pre-equilib\-rium particle emission and equilibrium evaporation, competing with fission and $\gamma$-ray emission. The pre-equilibrium emission manifests itself through the high-energy components of the emitted-particle spectra and can modify the excitation-energy and angular-momentum distributions of the daughter nuclei. Its contribution may therefore affect the population of low-lying states and the resulting residual-nucleus production cross sections. For heavy-ion-induced reactions, the pre-equilibrium processes can be significant \cite{Veselsky1997}.

One of the theoretical approaches commonly used to describe these processes combines the Hauser--Feshbach (HF) statistical model \cite{Hauser1952} with the contribution of the pre-equilibrium particle emission calculated within the Griffin exciton model \cite{Griffin1966}. Since the pre-equilibrium processes are more peripheral and are therefore closer to direct reactions, the pre-equilibrium particles generally have higher energies than the particles emitted during the equilibrium evaporation stage. Correspondingly, they may carry away larger energy and angular momenta.

In the present work, we apply this combined approach to analyze the production of the $^{249-251}$No isotopes in the $^{204}$Pb($^{48}$Ca,xn) $^{252-x}$No reaction, with particular emphasis on the production of $^{250g,m}$No in the ground and isomeric states. We investigate whether the inclusion of pre-equilibrium particle emission can account for the high-energy behavior of the measured production cross sections and examine its influence on the population of the ground and isomeric states. The experimental data also provide an opportunity to study the relationship between the angular momentum carried away by pre-equilibrium particles and the calculated isomeric ratio.

The formalism is implemented in the STAPRE-F code \cite{uhl1976stapre}, which has been modified for calculations at high excitation energies and large angular momenta and adapted for modern Fortran 95 compilers \cite{ISO1539-1:2004:DIS}. The angular-momentum range was extended up to $200\hbar$.

In the following, we briefly describe the theoretical formalism, discuss the choice of the model parameters, and compare the calculated results with the experimental data \cite{Tezekbayeva2022}.

%===============================================================================

\section{Theoretical approach}
\label{sec:theory}
%===============================================================================

%===============================================================================
\subsection{Hauser--Feshbach model}
\label{subsec:hf}
%===============================================================================

The central assumption of this model is the Bohr's independence hypothesis, which states that the formation and decay of a compound nucleus are independent processes. Once the statistical equilibrium \cite{Ribansky1973545} has been established, the compound nucleus decays through the emission of particles or $\gamma$ rays.

According to the HF model, the cross section for a reaction with entrance channel $a$ and exit channel $b$ is expressed in terms of the transmission coefficients as
\begin{equation}
	\sigma_{ab} \sim \frac{T_a T_b}{\sum \limits_i T_i},
\end{equation}
where the sum is taken over all open reaction channels.

In the present approach, we consider a sequence of emitted particles (neutrons in the reaction under study) competing with the emission of other light particles ($p$, $\alpha$, and $H$), $\gamma$ rays, and fission.

The transmission coefficients are related to the $S$ matrix calculated using a complex optical potential. In the present calculations, the heavy-ion transmission coefficients are evaluated within the smooth cutoff approximation suggested in Ref.~\cite{shidling2006fission,mancusi2010unified} as
\begin{eqnarray}
	T_l = \frac{1}{1+e^{(l-l_{\max})/\Delta l}},
	\label{eq:T_l}
\end{eqnarray}
where $\Delta l=1$, the cutoff angular momentum is given by
$l_{\max}\approx kR\sqrt{1-\frac{B_e}{E_{LS}}}$, the momentum is
$k=\sqrt{2E_{LS}\mu}$, $E_{LS}$ is the projectile energy, $\mu$ is the reduced mass of the projectile--target system, and
$R=r_0(A_t^{1/3}+A_p^{1/3})$, with $r_0=1.25$ fm, is the interaction radius. Here, $A_t$ and $A_p$ are the mass numbers of the target and projectile nuclei, respectively.

In the HF model, the fusion cross section is calculated as
\begin{equation}
	\sigma_{\mathrm{fus}}=
	\frac{\pi}{k^2}\sum_l^{l_{\max}}T_l(2l+1),
	\label{eq:sigma_fus}
\end{equation}
where $T_l$ is the transmission coefficient for the partial wave with angular momentum $l$.

Knowing $\sigma_{\mathrm{fus}}$ from experimental data, one can obtain the angular-momentum distribution of the compound nucleus from Eqs.~(\ref{eq:sigma_fus}) and~(\ref{eq:T_l}). In the present calculations, we used the experimental fusion-cross-section data for the $^{48}$Ca+$^{208}$Pb reaction \cite{Pacheco1992}.

The $(a,b)$ reaction cross section, where $a$ and $b$ denote the entrance and exit reaction channels, respectively, is given by
\begin{equation}
	\begin{aligned}
		\sigma_{ab}
		&= \pi \bar{\lambda}_a g_a
		\frac{\tilde{T}_a \tilde{T}_b}{\sum_{b'}\tilde{T}_{b'}} \\
		&= \sigma_{\mathrm{in}}
		\frac{\tilde{T}_b}{\sum_{b'}\tilde{T}_{b'}}
		= \sigma_{\mathrm{in}}
		\frac{\langle \Gamma_b \rangle}{\sum_{b'}\langle \Gamma_{b'} \rangle},
	\end{aligned}
\end{equation}
where the total inelastic cross section is
$\sigma_{\mathrm{in}}=\sum_b \sigma_{ab}$.

The total decay width of an excited compound nucleus is expressed as the sum of the decay widths for particle emission ($n$, $p$, $\alpha$, and $^{2}$H), fission, and $\gamma$-ray emission at each step of the evaporation cascade,
\begin{equation}
	\Gamma=\sum_i\Gamma_i+\Gamma_\gamma+\Gamma_f.
\end{equation}

All decay widths are calculated by taking into account transitions to both overlapping and isolated levels of the daughter nucleus,
\begin{equation}
	\Gamma_i(E,J,\wp)=
	\Gamma_i^{\mathrm{ov}}(E,J,\wp)+
	\Gamma_i^{\mathrm{iso}}(E,J,\wp),
\end{equation}
where $E$, $J$, and $\wp$ are the excitation energy, spin, and parity of the decaying nucleus, respectively. The decay width for the particle emission is written as

\begin{equation}
	\begin{aligned}
		\Gamma_i^{\mathrm{ov}}(E,J,\wp)
		&=\frac{1}{2\pi p_c(E,J,\wp)} \\
		&\times \sum_l \int_0^{E-B_i}
		G_l(\wp,\wp_d)\,
		T_l(E-B_i-E_d) \\
		&\times \sum_{j_d}
		\frac{2j_d+1}{2}
		\rho_d(E_d,J_d,\wp_d)\,\mathrm{d}E_d,
	\end{aligned}
\end{equation}
where $B_i$ is the binding energy of the $i^{th}$ particle ($i=n,p,\alpha,^{2}$H) in the nucleus, taken from Refs.~\cite{Huang2021AME2020I,Wang2021AME2020II}. The indices $c$ and $d$ denote the parent and daughter nuclei, respectively, whereas $G_l(\wp,\wp_d)$ accounts for parity conservation and $l$ is the orbital angular momentum of the emitted particle.

The decay width for the $\gamma$-ray channel is defined as
\begin{equation}
	\begin{aligned}
		\Gamma^{\mathrm{ov}}_{\gamma}(E, J, \wp)
		&= \frac{1}{2\pi p_c(E, J, \wp)}  \\
		&\times \sum_l \int_0^{E}
		G^M_l(\wp_c,\wp) e^{2l-1}\zeta_l  \\
		&\times \sum_{j_c}
		\frac{(2j_c+1)}{2}
		\rho_c(E-e, J_c, \wp_c)\,\mathrm{d}e,
	\end{aligned}
\end{equation}
where $\zeta_l$ is a normalization constant chosen such that the radiative decay width at the neutron binding energy reproduces the experimental radiative-width data, while $l$ and $M$ denote the multipolarity and type of the radiative transition, respectively.

To calculate the decay width in the fission channel, the Hill--Wheeler formula \cite{HillWheeler1953} is used,
\begin{equation}
	\begin{aligned}
		\Gamma^{\mathrm{ov}}_f(E, J, \wp)
		&= \frac{1}{2\pi p_c(E, J, \wp)}  \\
		&\times \sum_l \int_0^{E-B_f}
		\frac{\rho_f(E_f, J, \wp)\,\mathrm{d}E_f}
		{1+\exp\left[2\pi(B_f-E_f)/\hbar\omega\right]} .
	\end{aligned}
\end{equation}
where $\hbar\omega$ is the curvature of the fission barrier (oscillator energy), $B_f$ is the fission-barrier height, $E_f$ is the energy of the fission fragment, and $\rho_f(E,J,\wp)$ is the level density at the saddle point of the fission barrier.

In the case of a double-humped fission barrier, the total fission width is calculated assuming weak coupling between the two classes of excited states and a relatively low probability of the particle and $\gamma$-ray emission from excited states in the second potential well.

For the level-density calculations, we use the back-shifted Fermi-gas model \cite{Dilg1973} with the ratio of the effective and rigid-body moments of inertia chosen as
${\cal F}/{\cal F}_{\mathrm{rb}}=0.5$.
The level density is given by
\begin{equation}
	\rho(E, J, \wp) =
	\frac{(2J+1)\exp\left(2\sqrt{aU}\right)}
	{24\sqrt{2}\sigma^3 a^{1/4}E^{5/4}}
	\exp\left[-\frac{(J+1/2)^2}{2\sigma^2}\right],
\end{equation}
where $\sigma$ is the spin-cutoff parameter,
$\sigma^2={\cal F}\frac{\sqrt{E/a}}{\hbar^2}$,
$a$ is the level-density parameter,
${\cal F}$ is the moment of inertia of the nucleus, and
$U=E-\Delta$ is the effective excitation energy corrected by the back-shift parameter $\Delta$.

\subsection{Exciton model}
\label{subsec:exciton}
%===============================================================================

In the exciton model, the state of the system is characterized by the excitation energy of the nucleus, $E$, and the number of excitons, $n=p+h$, where $p$ is the number of particles above the Fermi energy and $h$ is the number of holes below it. The transition rules between different exciton states are $\Delta n = 0,\pm 2$. An exciton state can also decay through particle emission.

In contrast to the HF model, where the decay of an excited state occurs independently of its formation, in the exciton model the excited state is genetically related to the initial state, and the relaxation toward the equilibrium state is described by the time-dependent master equation \cite{KalbachCline1972,Ribansky1973545},
\begin{equation}
	\begin{aligned}
		\frac{\mathrm{d}P(n,t)}{\mathrm{d}t}
		&= P_{n-2}(t)\lambda_{n-2}^+(E)
		+ P_{n+2}(t)\lambda_{n+2}^-(E) \\
		&+ P_n(t)\lambda_n^0(E) \\
		&- P_n(t)\left[\lambda_n^+(E)+\lambda_n^-(E)+W_{n}(E)\right]
	\end{aligned}
	\label{eq:Master}
\end{equation}
where $P_n(t)$ is the probability of finding the system in the exciton state $n$ at time $t$, $\lambda_n^{\pm}(E)$ is the probability of an intranuclear transition per unit time, and $W_{n}(E)$ describes the pre-equilibrium particle emission.

The initial condition is
\begin{equation}
	P(n,0)=\delta_{n,n_0}=\delta_{p,p_0}\delta_{h,h_0},
\end{equation}
where $(p_0,h_0)$ is the initial exciton configuration and represents a free parameter of the model.

The rate of internal transitions is determined by the squared matrix element
$\langle |M|^2 \rangle$ as
\begin{equation}
	\lambda_{0,\pm}=
	\frac{2\pi}{\hbar}\langle |M|^2 \rangle\omega^{0,\pm}(E),
\end{equation}
where $\omega^{0,\pm}$ is the density of excited states of the system with excitation energy $E$ \cite{williams1970intermediate},
\begin{equation}
	\begin{aligned}
		\omega^+ &= \frac{g(gE-A_{ph})^2}{n+1},\\
		\omega^0 &= g(gE-A_{ph})\frac{3n-2}{4},\\
		\omega^- &= gph(n-2).
	\end{aligned}
\end{equation}

Here, $A_{ph}$ is the Pauli-principle correction. These expressions include the parameter $g$, the single-particle level density, which is related to the level-density parameter in the Fermi-gas model through
$g=6a/\pi^2$. Thus, the parameter $g$ enters the definition of the principal quantities of both the exciton and HF models and serves as their common parameter.

The matrix elements of the transition rates between the exciton states are taken in the standard form \cite{KalbachCline1972}
\begin{equation}
	\langle |M|^2 \rangle =
	\frac{k n}{A^3 E}
	\begin{cases}
		\left(\dfrac{E}{7n}\right)^{1/2}
		\left(\dfrac{E}{2n}\right)^{1/2},
		& \text{for } \dfrac{E}{n}<2~\text{MeV}, \\[2mm]
		
		\left(\dfrac{E}{7n}\right)^{1/2},
		& \text{for } 2 \leq \dfrac{E}{n}<7~\text{MeV}, \\[2mm]
		
		1,
		& \text{for } 7 \leq \dfrac{E}{n}\leq15~\text{MeV}, \\[2mm]
		
		\left(\dfrac{E}{15n}\right)^{1/2},
		& \text{for } \dfrac{E}{n}>15~\text{MeV}.
	\end{cases}
	\label{eq:MatElem}
\end{equation}
where $k=135$ MeV$^3$.

The fusion cross section $\sigma_{\mathrm{fus}}$ in Eq.~(\ref{eq:sigma_fus})
consists of the contributions from the pre-equilibrium cross section
$\sigma^{\mathrm{PE}}$ and the compound-nucleus formation cross section
$\sigma_c$,
\begin{equation}
	\sigma_{\mathrm{fus}}=\sigma^{\mathrm{PE}}+\sigma_c,
	\label{eq:pe_eq}
\end{equation}
where $\sigma^{\mathrm{PE}}$ is defined through the pre-equilibrium fraction
$p^{\mathrm{PE}}$ as
$\sigma^{\mathrm{PE}}=p^{\mathrm{PE}}\sigma_{\mathrm{fus}}$.
The cross section $\sigma^{\mathrm{PE}}$ is calculated within the exciton model by solving the time-dependent master equation~(\ref{eq:Master}) \cite{KalbachCline1972} until the equilibrium conditions \cite{Ribansky1973545} are satisfied.

The particle spectra are obtained from all the exciton states populated during the relaxation toward the equilibrium state. The emission rate of a particle of type $b$ from the exciton state $n$ is given by
\begin{equation}
	W(n,E)=\sum_bW_b(n,E)=
	\sum_b\int_0^{E-B_b}
	\lambda_b(n,\varepsilon_b)\,\mathrm{d}\varepsilon_b,
	\label{eq:partemissionprob}
\end{equation}
where the summation is performed over all open particle-emission channels, and
$\lambda_b(n,\varepsilon_b)$ is the emission rate of particle $b$ with energy
$\varepsilon_b$,
\begin{equation}
	\lambda_b(n,\varepsilon_b)=
	\frac{2s_b+1}{\pi^2\hbar^3}
	\mu_b\varepsilon_b\sigma_{\mathrm{inv}}(\varepsilon_b)
	R_b(p)
	\frac{\omega(p-n_b,h,U_b)}{\omega(p,h,E)}.
	\label{eq:emissionrate}
\end{equation}

Here, $n_b$ is the number of nucleons in particle $b$, while $s_b$ and $\mu_b$ are the spin and reduced mass of the emitted particle, respectively. The factor $R_b(p)$ accounts for the charge conservation. The quantity $\sigma_{\mathrm{inv}}$ is the inverse cross section, in which the compound nucleus (here $^{252}$No) is formed through the fusion of particle $b$ with the residual nucleus, and is calculated using Eq.~(\ref{eq:sigma_fus}).

The exciton-state level density $\omega(p,h,E)$ depends on the exciton configuration $(p,h)$ and the excitation energy, and is determined by the single-particle level density $g$ \cite{ignatyuk1978dependence},
\begin{equation}
	\omega_{p,h}(E,M)=
	\frac{g(gU)^{p+h-1}}
	{2^{p+h}h!p!(p+h-1)!}
	C^{(p+h-M/m)/2}_{p+h},
	\label{eq:leveldensexciton}
\end{equation}
where the effective excitation energy is corrected according to
$U=E-\Delta$, and $M$ is the angular momentum.

Note that the transmission coefficients and level-density parameters are used consistently in both the HF and exciton models. It is assumed that the penetrability coefficients calculated within the optical model are applicable to nuclei in both equilibrium and non-equilibrium states. It is also assumed that, in the exciton model, the momentum distribution is the same as that in the HF model. We performed our calculations with two different angular momentum distributions (see discussion below).

After determining the contribution of $\sigma^{\mathrm{PE}}$ to the fusion cross section and the pre-equilibrium fraction,
$p^{\mathrm{PE}}=\sigma^{\mathrm{PE}}/\sigma_{\mathrm{fus}}$,
the remaining part of the fusion cross section corresponds to evaporation processes and is used in the HF-model calculations of the momentum distributions of the daughter nuclei. In the present work, we consider the sequential evaporation of neutrons in the $^{204}$Pb($^{48}$Ca,xn)$^{252-x}$No reaction. At each stage of the de-excitation cascade, the emission of light particles ($n$, $p$, $\alpha$, and $^{2}$H) competes with $\gamma$-ray emission and fission.

\subsection{Input parameters}
\label{subsec:input_parameters}
%===============================================================================

The free parameters of the models described above include the transmission coefficients, the level-density parameters for all nuclei produced in the reaction, $a$, $\Delta$, ${\cal F}/{\cal F}_{\mathrm{rb}}$, the fission-barrier parameters, namely the fission-barrier heights $B_f$, their curvatures $\hbar\omega$, the level density at the saddle point of the fission barrier, $\rho_f(E,J,\wp)$, and the initial exciton configuration $(p_0,h_0)$.

The level densities are calculated using the back-shifted Fermi-gas model with the back-shift parameter taken from the systematics of Dilg \cite{Dilg1973}. We choose parameter values typical for heavy nuclei, namely
$a=25$ MeV$^{-1}$,
$\Delta=0.75$,
and
${\cal F}/{\cal F}_{\mathrm{rb}}=0.5$.
It should be noted that a more accurate determination of these parameters would require additional experimental data. In the present work, our primary objective is to demonstrate the role of pre-equilibrium processes in the production of nobelium isotopes, and these parameter values are sufficiently accurate for this purpose. Their variation does not significantly affect the energy dependence of the production cross sections.

For the calculation of the transmission coefficients in the $^{48}$Ca-induced reaction, the parameter
$\Delta l=10$
in Eq.~(\ref{eq:T_l}) is used (see Ref.~\cite{mancusi2010unified}).

In our calculations, the fission-barrier heights
$B_{\mathrm{fis}}$
(which may include one or more barriers, $i>1$)
and their curvatures
$\hbar\omega_i$
are taken from the estimates of Ref.~\cite{Moller2015}.
The adopted parameter values are listed in Table~\ref{tab:barriersparam}. The level-density parameter at the fission barrier,
$a=30$ MeV$^{-1}$,
and the back-shift parameter,
$\Delta=0.75$ MeV,
are used for all fissioning No isotopes.

\begin{table}[t]
	\centering
	\caption{Parameters of the fission-barrier heights $B_{f_i}$ and curvatures $\hbar\omega_i$, where $i$ denotes the barrier number.}
	\label{tab:barriersparam}
	\begin{tabular}{lcccc}
		\hline
		& $B_{f_1}$, MeV & $\hbar\omega_1$, MeV
		& $B_{f_2}$, MeV & $\hbar\omega_2$, MeV \\
		\hline
		$^{251}$No & 10.75 & 3.6 & 6.52 & 3.6 \\
		$^{250}$No & 6.5  & 3.6 & --   & --  \\
		$^{249}$No & 5.5  & 3.6 & --   & --  \\
		\hline
	\end{tabular}
\end{table}

The $\gamma$-decay schemes of the $^{250,251}$No isotopes are taken from Ref.~\cite{LopezMartens2026}. It should be noted that the available information on the branching ratios in the $\gamma$-decay schemes is incomplete, although these ratios strongly affect the calculated isomeric ratio. Therefore, we compare our results with the experimental production cross sections of $^{250}$No in the ground and isomeric states.

The radiative width at the neutron binding energy is taken from the RIPL systematics for heavy nuclei \cite{capote2009ripl} and is approximately 40 meV. The initial exciton configuration in the exciton model is fitted to reproduce the experimental data on the production of nobelium isotopes, yielding
$(p_0,h_0)=(4,4)$.

Binding energies are taken from Refs.~\cite{Huang2021AME2020I,Wang2021AME2020II}.

\section{Results and discussion}
\label{sec:results}
%===============================================================================

\begin{figure}[t]
	\centering
	\includegraphics[width=0.5\textwidth]{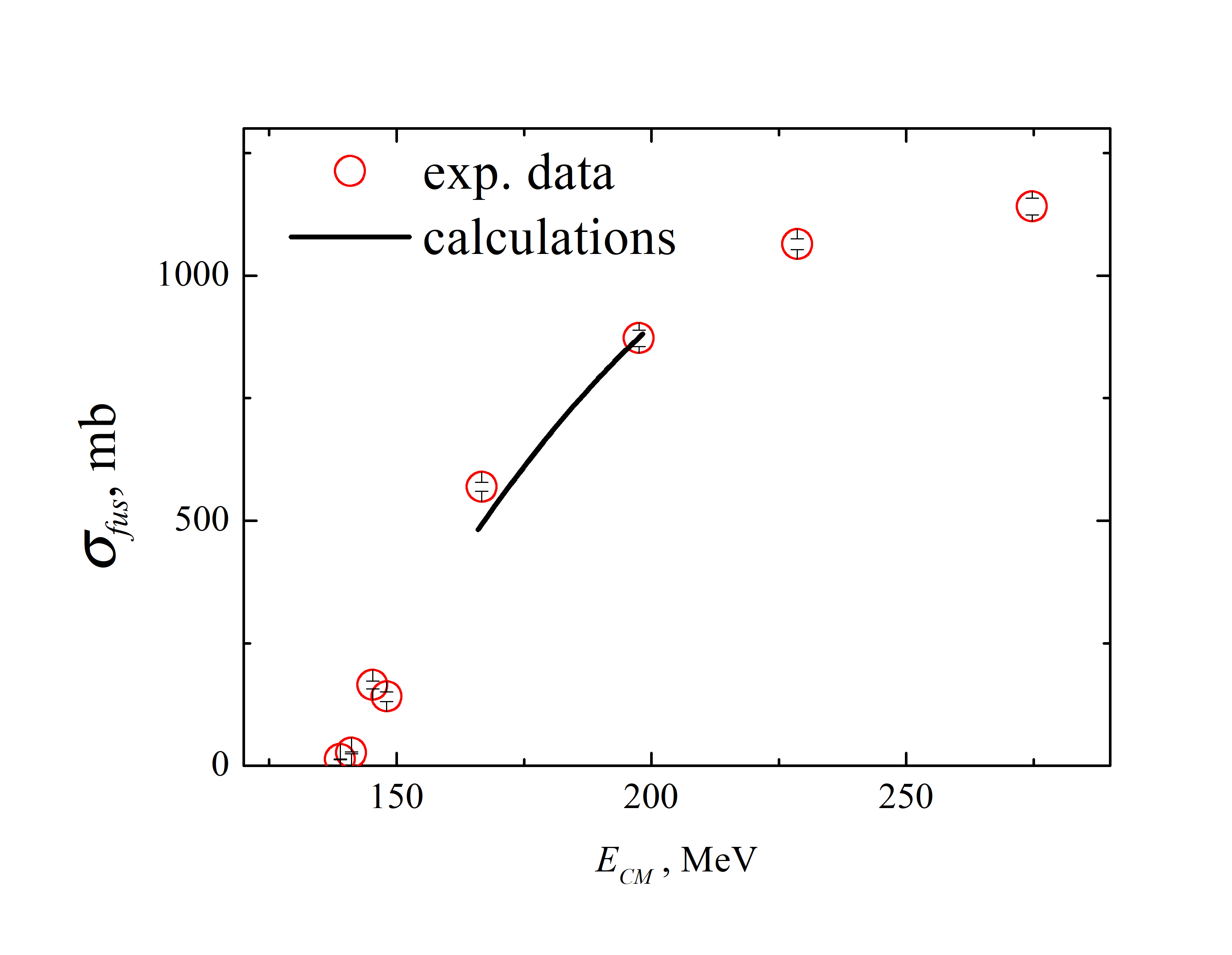}
	\caption{Fusion cross sections used in the present calculations together with the experimental data from Ref.~\cite{Pacheco1992}.}
	\label{fig:sigfus}
\end{figure}

Figure~\ref{fig:sigfus} presents the fusion cross sections used in our calculations. They were obtained by interpolating the experimental data from Ref.~\cite{Pacheco1992} for the $^{48}$Ca+$^{204}$Pb fusion reaction and correcting them by the geometric factor
\[
(A_{204\mathrm{Pb}}^{1/3}+A_{48\mathrm{Ca}}^{1/3})/
(A_{208\mathrm{Pb}}^{1/3}+A_{48\mathrm{Ca}}^{1/3}).
\]

%-------------------------------------------------------------------------------
%-------------------------------------------------------------------------------

\begin{figure}[t]
	\centering
	\includegraphics[width=0.5\textwidth]{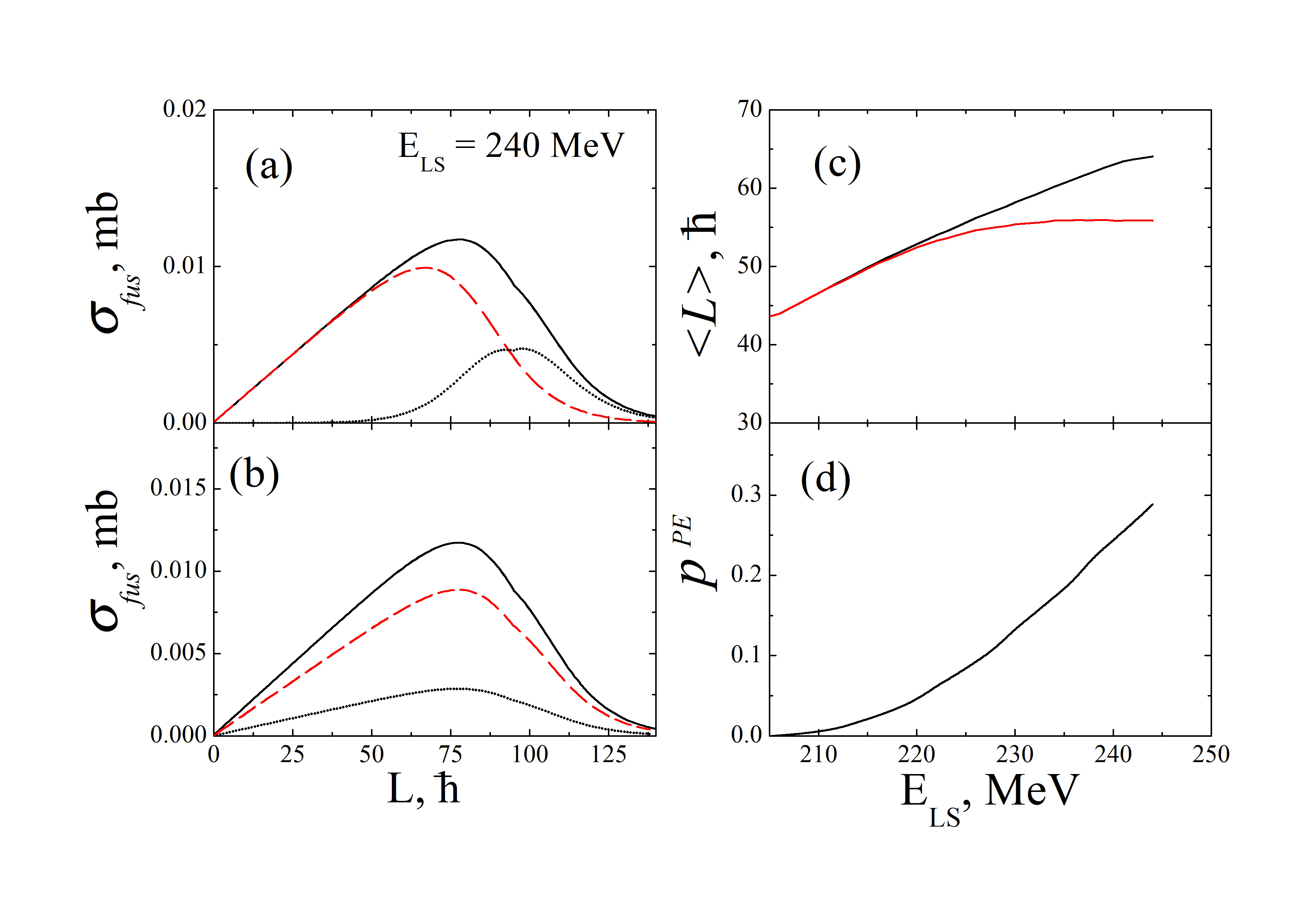}
	\caption{
		Panels (a) and (c) show the angular-momentum distribution of $^{252}$No (solid line), together with its equilibrium (dashed line) and pre-equilibrium (dotted line) components calculated for cases (i) and (ii), respectively. Panel (b) presents the average angular momentum of the equilibrium component calculated for cases (i) (red line) and (ii) (black line). Panel (d) shows the calculated pre-equilibrium fraction.}
	\label{fig:2}
\end{figure}

Using Eqs.~(\ref{eq:T_l}) and~(\ref{eq:sigma_fus}) together with the experimental fusion cross sections, we obtain the angular-momentum distribution of the first compound nucleus. The resulting distribution is shown by the solid black line in Figs.~\ref{fig:2}(a) and~\ref{fig:2}(c).

The angular-momentum distributions of $\sigma^{\mathrm{PE}}$ and $\sigma_C$ are calculated under two assumptions:
(i) the pre-equilibrium particles carry away the maximum angular-momentum [Fig.~\ref{fig:2}(a)], and
(ii) the pre-equilibrium emission does not modify the angular-momentum distribution [Fig.~\ref{fig:2}(c)].
The angular-momentum distributions corresponding to the equilibrium and pre-equilibrium contributions in Eq.~(\ref{eq:pe_eq}) are shown by the dashed and dotted curves, respectively.

The change in the angular-momentum distribution in case (i) results in a reduction of the average angular momentum associated with $\sigma_C$ (see Fig.~\ref{fig:2}(b)). As shown in Fig.~\ref{fig:2}(b), high-angular-momentum states up to
$L_{\max}\sim120\hbar$
are populated in this reaction, whereas the corresponding change in the average angular momentum remains relatively small. Consequently, no noticeable difference is observed in the calculated production of the $^{249-251}$No isotopes between cases (i) and (ii) when pre-equilibrium emission is taken into account.

Figure~\ref{fig:2}(d) shows the energy dependence of the pre-equilibri\-um fraction
$p^{\mathrm{PE}}$
in the fusion cross section. The production of the $^{249-251}$No isotopes strongly depends on this fraction.

%------------------------------------------------------------------------------
%------------------------------------------------------------------------------

\begin{figure}[t]
	\centering
	\includegraphics[width=0.5\textwidth]{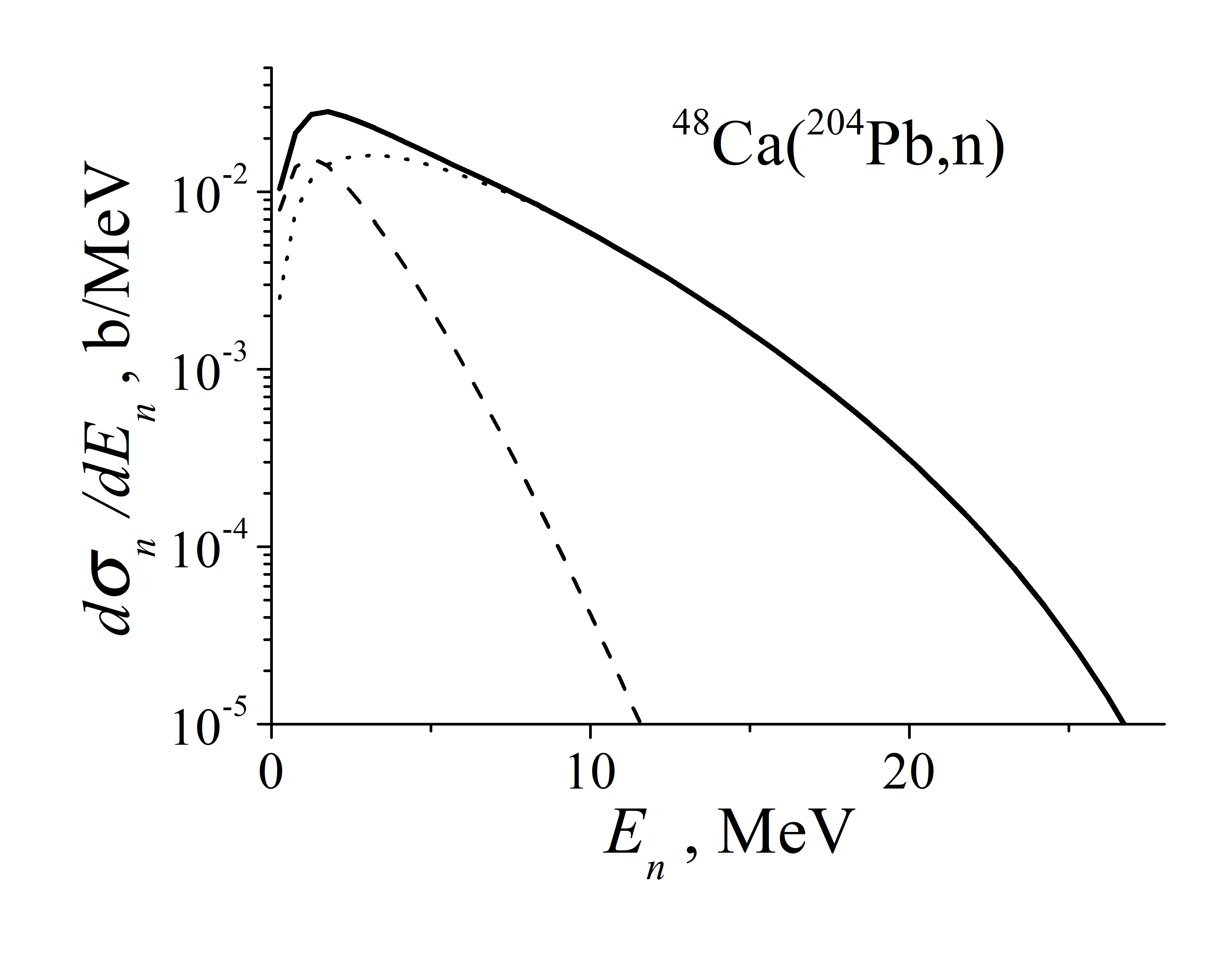}
	\caption{
		Energy spectra of neutrons emitted from $^{252}$No through equilibrium evaporation (dashed line), pre-equilibrium emission (dotted line), and the total neutron spectrum (solid line) at the incident energy
		$E=240$ MeV.}
	\label{fig:3}
\end{figure}

Figure~\ref{fig:3} shows the neutron energy spectra for $^{252}$No formed in the
$^{48}$Ca+$^{204}$Pb fusion reaction. The equilibrium and pre-equilibrium contributions to the neutron spectra are shown for the incident energy
$E_{\mathrm{inc}}=240$
MeV. A difference of several orders of magnitude is observed in the high-energy part of the spectra due to the pre-equilibrium contribution. As a result, the low-lying states of the daughter nucleus $^{250}$No are populated more efficiently.

%------------------------------------------------------------------------------
%------------------------------------------------------------------------------

At this energy, the pre-equilibrium fraction is 0.24 (24\%). The fusion cross section includes contributions from pre-equilibri\-um particle emission, equilibrium particle evaporation, and fission. The fission cross section of $^{252}$No amounts to approximately 90\% of $\sigma_C$ and is treated as an equilibrium process. 

Consequently, at the incident energy of 240 MeV, pre-equilib\-rium
processes reduce the compound-nucleus contribution $\sigma_C$ to the
fusion cross section [Eq.~(\ref{eq:pe_eq})] to approximately 70\%,
whereas the particle-emission contribution increases from about 10\%
to 30\%, i.e., by a factor of three. As a result, the equilibrium and
pre-equilibrium neutron yields become comparable, making the influence
of pre-equilibrium emission particularly pronounced for the production
of all No isotopes considered.

The importance of pre-equilibrium emission increases with increasing
incident energy. At 220 MeV, the pre-equilibrium fraction is only about
5\%, and the corresponding neutron yield is approximately five times
smaller than that from equilibrium evaporation. At 230 MeV, the
difference decreases to about a factor of two, whereas at 240 MeV the
equilibrium and pre-equilibrium neutron yields become comparable.
Thus, even a relatively small pre-equilibrium fraction can substantially
modify the neutron spectrum, leading to a significant enhancement of
the population of low-lying states in the daughter nuclei and,
consequently, to increased production cross sections of the residual
No isotopes.

\begin{figure}[t]
	\centering
	\includegraphics[width=\linewidth]{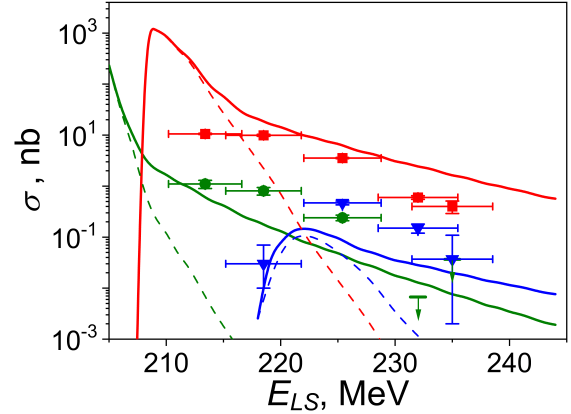}
	\caption{Calculated production cross sections of the $^{249-251}$No isotopes in the $^{204}$Pb($^{48}$Ca,xn)$^{252-x}$No reaction obtained with $k=0$ MeV$^3$ (dashed curves) and $k=135$ MeV$^3$ (solid curves), together with the experimental data from Ref.~\cite{Tezekbayeva2022}. The experimental data were obtained at mid-target energies $E_{1/2}$ between 200 and 240 MeV. Green, red, and blue curves correspond to the production of the $^{251}$No, $^{250}$No, and $^{249}$No isotopes, respectively, through sequential neutron emission from $^{252}$No. The experimental data are shown by green circles for the 1$n$ channel ($^{251}$No), with cross-section limits indicated for the cross sections at 232 and 235 MeV, red squares for the 2$n$ channel ($^{250}$No), and blue triangles for the 3$n$ channel ($^{249}$No).}
	\label{fig:4}
\end{figure}

Figure~\ref{fig:4} compares the calculated production cross sections of the $^{249-251}$No isotopes with the experimental data from Refs.~\cite{Tezekbayeva2021,Tezekbayeva2022}. The experimental data were obtained at the mid-target energies $E_{1/2}$
between 213 and 240 MeV \cite{Tezekbayeva2021}. The dashed curves correspond to calculations without pre-equilibrium emission ($k=0$ MeV$^3$ in Eq.~(\ref{eq:MatElem})), whereas the solid curves include the contribution of pre-equilibrium nucleon emission ($k=135$ MeV$^3$).

Above the reaction threshold, a pronounced peak associated with neutron evaporation (equilibrium emission) from the residual No isotopes is observed. Similar structures were reported previously in Ref.~\cite{Tezekbayeva2022}. In addition, high-energy tails originating from pre-equilibrium emission are clearly visible.

The results demonstrate that pre-equilibrium processes play an essential role in the production of the residual nuclei. Taking pre-equilibrium emission into account allows the high-energy behavior of the measured production cross sections of the No isotopes to be reproduced.

\begin{figure}[t]
	\centering
	\includegraphics[width=\linewidth]{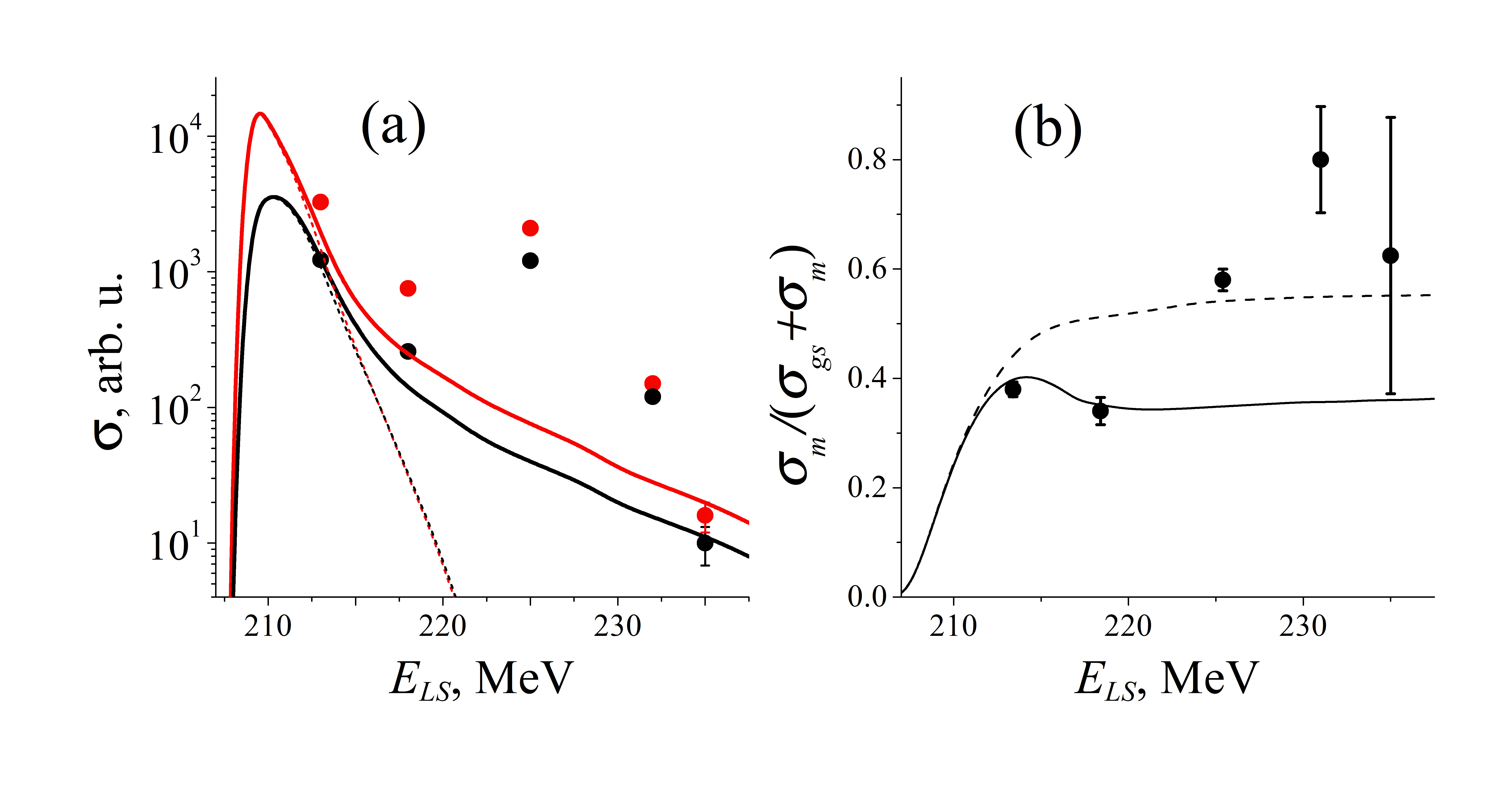}
	\caption{(a) Production cross sections of $^{250}$No in the ground (solid black curve) and isomeric (solid red curve) states calculated with $k=135$ MeV$^3$. The corresponding calculations with $k=0$ MeV$^3$ are shown by the dashed black and red curves, respectively. (b) Isomeric ratio for $^{250}$No. The symbols represent the experimental data from Ref.~\cite{Tezekbayeva2022}; the solid and dashed curves correspond to calculations with $k=135$ and $k=0$ MeV$^3$, respectively.}
	\label{fig:5}
\end{figure}

Fig.~\ref{fig:5} shows the calculated production cross sections of $^{250}$No in the ground and isomeric states together with the corresponding isomeric ratios at incident $^{48}$Ca energies from 213 to 240 MeV. The solid curves include the contribution of pre-equilibrium emission, whereas the dashed curves correspond to pure evaporation processes.

It should be noted that the absolute values of the isomeric ratios shown in Fig.~\ref{fig:5}(b) strongly depend on the branching ratios in the $\gamma$-decay scheme of $^{250}$No. As shown in Refs.~\cite{fotina1998analysis,parfenova1996study}, the isomeric ratio is highly sensitive to variations in the level densities of positive- and negative-parity states. These variations are related to the shell structure of the residual No isotopes and may be particularly important for low-lying states. Because the available information on the $\gamma$-decay scheme and branching ratios is incomplete, only qualitative conclusions regarding the calculated isomeric ratios can be drawn.

At the same time, the absolute production cross sections of $^{250}$No in the ground and isomeric states provide important information. Figure~\ref{fig:5}(a) shows the incident-energy dependence of the production of $^{250}$No in the ground (solid black curve) and isomeric (solid red curve) states. The calculations performed with $k=135$ MeV$^3$ in Eq.~(\ref{eq:MatElem}) are shown by the solid curves, whereas those obtained with $k=0$ MeV$^3$ are shown by the dashed curves. The influence of pre-equilibrium processes is similar to that discussed above: calculations including only equilibrium evaporation underestimate the experimental data in the high-energy region \cite{Tezekbayeva2022}.

%-------------------------------------------------------------------------------

The inclusion of pre-equilibrium emission significantly improves the agreement between the calculated and experimental production cross sections of $^{250g,m}$No.

\section{Conclusions}
\label{sec:conclusions}
%===============================================================================

The combined Hauser--Feshbach and exciton models were applied to describe the experimental data \cite{Tezekbayeva2022} on the production of the $^{249-251}$No isotopes in the $^{204}$Pb($^{48}$Ca,~xn)$^{252-x}$No reaction at $^{48}$Ca incident energies from 200 to 240 MeV.

It was found that including pre-equilibrium processes provides a significant improvement in the qualitative and quantitative description of the experimental data. The contribution of pre-equilibrium particle emission leads to changes in the production cross sections of the No isotopes by several orders of magnitude in the high-energy region.

These results demonstrate that pre-equilibrium emission plays an important role in heavy-ion-induced fusion reactions and should therefore be taken into account in the analysis of experimental production cross sections.

%===============================================================================

%\begin{acknowledgments}
	\section*{Acknowledgments}

This work was supported by the Science Committee of the Ministry of
Science and Higher Education of the Republic of Kazakhstan
(Grant No. AP27510285).

\bibliographystyle{elsarticle-num-names} 
	\bibliography{all}
	
	\end{document}